# Effect of Statistical Fluctuation in Monte Carlo Based Photon Beam Dose Calculation on Gamma Index Evaluation

Yan Jiang Graves[1,2], Xun Jia[1], Steve B. Jiang[1]

[1]Center for Advanced Radiotherapy Technologies and Department of Radiation Medicine and Applied Sciences, University of California San Diego, La Jolla, CA 92037-0843, USA
[2]Department of Physics, University of California San Diego, La Jolla, CA 92093, USA

E-mails: sbjiang@ucsd.edu, xunjia@ucsd.edu

The γ-index test has been commonly adopted to quantify the degree of agreement between a reference dose distribution and an evaluation dose distribution. Monte Carlo (MC) simulation has been widely used for the radiotherapy dose calculation for both clinical and research purposes. The goal of this work is to investigate both theoretically and experimentally the impact of the MC statistical fluctuation on the γ-index test when the fluctuation exists in the reference, the evaluation, or both dose distributions. To the first order approximation, we theoretically demonstrated in a simplified model that the statistical fluctuation tends to overestimate γ-index values when existing in the reference dose distribution and underestimate γ-index values when existing in the evaluation dose distribution given the original γ-index is relatively large for the statistical fluctuation. Our numerical experiments using realistic clinical photon radiation therapy cases have shown that 1) when performing a γ-index test between an MC reference dose and a non-MC evaluation dose, the average γ-index is overestimated and the gamma passing rate decreases with the increase of the statistical noise level in the reference dose; 2) when performing a γ-index test between a non-MC reference dose and an MC evaluation dose, the average γ-index is underestimated when they are within the clinically relevant range and the gamma passing rate increases with the increase of the statistical noise level in the evaluation dose; 3) when performing a γ-index test between an MC reference dose and an MC evaluation dose, the gamma passing rate is overestimated due to the statistical noise in the evaluation dose and underestimated due to the statistical noise in the reference dose. We conclude that the γ-index test should be used with caution when comparing dose distributions computed with Monte Carlo simulation.

**1. Introduction**

Dose distribution comparison is a frequently performed task in radiotherapy, where the degree of agreement between an evaluation dose distribution and a reference one is established using some quantitative metrics. For example, in a typical patient-specific quality assurance procedure for a radiotherapy treatment, a treatment plan is delivered to a phantom before the actual treatment, and the measured dose distribution is compared with the calculated dose distribution by the treatment planning system. Over the years, several dose comparison methods have been developed, including the quantitative dose difference test, the distance-to-agreement (DTA) test (Vandyk *et al.*, 1993; Harms *et al.*, 1998), the composite analysis for both dose difference and DTA (Shiu *et al.*, 1992; Cheng *et al.*, 1996; Harms *et al.*, 1998), the γ-index test (Low *et al.*, 1998; Depuydt *et al.*, 2002; Low and Dempsey, 2003; Bakai *et al.*, 2003; Stock *et al.*, 2005; Wendling *et al.*, 2007; Ju *et al.*, 2008; Chen *et al.*, 2009; Yuan and Chen, 2010; Low, 2010; Gu *et al.*, 2011b), and the test based on the maximum allowed dose difference (MADD) (Jiang *et al.*, 2006). Among these, the γ-index test is the one most commonly used. This method combines quantifications of dose differences between two dose distributions in both the dose domain and the spatial domain. This allows for the toleration of spatial shifts when comparing the two dose distributions, which is clinically acceptable, and avoids an exaggeration of dose differences in the area of a high dose gradient. Moreover, the γ-index test is quantitatively comprehensible. Based on the user specified criteria, e.g. 3% for the dose difference criterion and 3 mm for the DTA criterion (3%-3mm), the user can judge how good the agreement is based on the value of the γ-index. The smaller the γ-index value is, the closer the two dose distributions are.

In many contexts, the γ-index test is used to evaluate the agreement between two dose distributions, where one, or both of them, is calculated by Monte Carlo (MC) simulations. For instance, because of the high accuracy of the MC method it is desirable to verify the dose distribution calculated by a treatment plan system by comparing it with another one independently calculated using an MC method (Calvo *et al.*, 2012). This situation is becoming more and more common as MC dose calculations become more efficient with novel algorithms and hardware developments (Jia *et al.*, 2010; Jia *et al.*, 2011; Hissoiny *et al.*, 2011; Wang *et al.*, 2011). It is also common, when developing new dose calculation algorithms, to consider the MC dose as the golden standard and test the accuracy of the new algorithms against it via γ-index tests (Jelen and Alber, 2007; Gu *et al.*, 2011a; Hissoiny *et al.*, 2011).

However, MC is a statistical method and the statistical fluctuation is unavoidable in the resulting dose distributions. This fluctuation may have non-negligible impacts on the γ-index values and hence lead to biased conclusions from the γ-index test (Low and Dempsey, 2003; Low, 2010). This fact is easily understood from the graphical interpretations of the γ-index (Ju *et al.*, 2008). Let us consider a simple case where two 1D dose distributions are compared as in Figure 1(a). Suppose we plot the two dose distributions, $\overline{D_r}$ and $\overline{D_e}$ with normalized coordinates $r/\Delta r$ and $D/\Delta D$, where $\overline{D_r}$ and $\overline{D_e}$ are the normalized reference dose distribution and the normalized evaluation dose distribution respectively; $\Delta r$ and $\Delta D$ are DTA and dose-difference criteria, respectively. It has been shown that the γ-index value at a coordinate $r$, denoted as $\gamma_0$, is simply the minimum Euclidian distance from the point $O_r$ to the evaluation dose distribution, which is graphically represented by a circle centered at $O_r$ such that the evaluation dose distribution is tangent to it. Figure 1(b) illustrates an example of how the γ-index value changes due to the MC statistical fluctuations in the evaluation dose. Suppose with fluctuations the evaluation dose distribution $\overline{D_e}$ becomes $\overline{D_e'}$, the new γ-index value $\gamma_0'$ could be different from the original one. Similarly, Figure 1(c) illustrates an example where the γ-index value is affected by the MC statistical fluctuations in the reference dose. Considering these scenarios, a few key questions need to be answered before comparing two dose distributions where an MC dose is involved: while the γ-index value apparently depends on a specific random realization of the dose distributions, what is the impact on average? How significant is this impact on, the clinically more important quantity, gamma passing rate? Also, since the γ-index value is not symmetric with respect to the two distributions (Low and Dempsey, 2003; Low, 2010), we have one more question: Is the impact different when the statistical fluctuation exists in the reference dose, in the evaluation dose, or in both of them?

The purpose of this paper is to systematically investigate the impact of the MC statistical fluctuation on the γ-index test. Specifically, we will demonstrate in a simplified model that, to the first order approximation, statistical fluctuation in the reference dose tends to overestimate γ-index values, while that in the evaluation dose tends to underestimate γ-index values when they are within the clinically relevant range. We also demonstrate these effects using one prostate and one head-and-neck (HN) clinical photon radiation therapy cases.

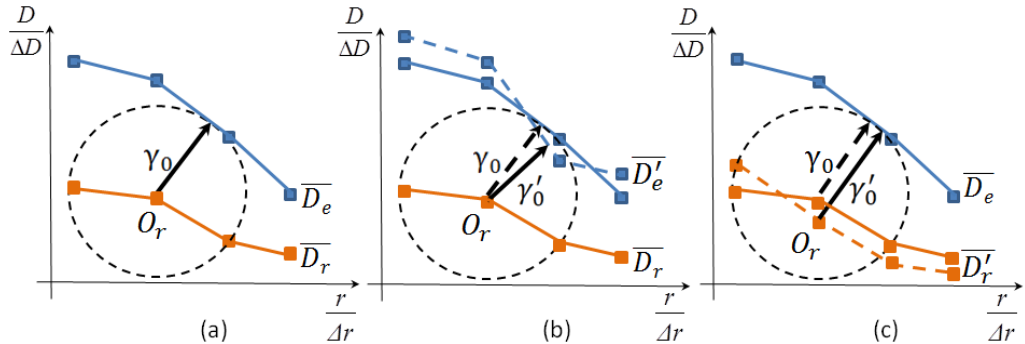

**Figure 1 (a).** Graphical interpretation of the γ-index in one-dimension. **(b).** An example demonstrating how the γ-index value changes due to MC statistical fluctuations in the evaluation dose. **(c).** An example demonstrating how the γ-index value changes due to MC statistical fluctuations in the reference dose.

**2. Methods and Materials**

*2.1 γ-index test*

Let us consider two dose distributions to be compared, $D_r$ and $D_e$. Mathematically, the γ-index at a comparison point $(r_r, D_r(r_r))$ is defined as

$$\gamma(r_r) = \min_{r_e}\left\{\sqrt{\frac{|r_e-r_r|^2}{\Delta r^2} + \frac{|D_e(r_e)-D_r(r_r)|^2}{\Delta D^2}}\right\}. \quad (1)$$

Here $D_r(r_r)$ is the reference dose distribution at position $r_r$ and $D_e(r_e)$ is the evaluation dose distribution at position $r_e$. $\Delta D$ is dose difference criterion and $\Delta r$ is the DTA criterion. If the γ-index is equal to or less than 1, the dose at the spatial point $r_r$ is considered to pass the test.

The graphical interpretation of the γ-index has been established previously (Ju *et al.*, 2008) and illustrated in Figure 1(a). Specifically, the γ-index is defined as the minimum Euclidian distance from the point $O_r$ to the evaluation dose curve. In practice, the evaluation dose is usually defined at discrete spatial locations. To the first order approximation, Ju *et al* assumed a linear interpolation of the dose values between neighboring spatial points as shown in Figure 1, and developed a simple and efficient γ-index calculation algorithm. This method is widely used nowadays (i.e. Gu *et al.*, 2011b). In this study, we keep this linear interpolation and our analysis in next section is built based on this implementation.

*2.2. Theoretical analysis of MC statistical fluctuation effects on the γ-index test*

The γ-index value for a particular random realization of dose distributions is essentially a random variable. It is more meaningful to investigate how the γ-index test is affected on average, *i.e.* the average impact over all the random dose realizations. In this section, we outline a theoretical calculation in a simplified 1D model for the mean γ-index value when there exist MC statistical fluctuations in the dose distributions. Theoretical analysis based on this simplified model is intended to offer some theoretical insights rather than a strict theoretical proof. We then validate our theoretical conclusions for 3D cases using real clinical data.

*2.2.1 MC statistical fluctuations in the reference dose*

In Figure 2, $O_{e1}O_{e2}$ is the line segment of the evaluation dose curve, $O_r$ is the reference point, $\gamma_0$ is the original γ-index value and $t$ parameterizes the deviation due to the statistical fluctuation. In this simplified model, the new γ-index value is always the minimum Euclidian distance from the point $O_r$ to the line segment $O_{e1}O_{e2}$ of the evaluation dose curve, not to the other line segment of the evaluation dose curve. Here we first introduce the concept of signed-gamma index $\tilde{\gamma}$, such that its magnitude is the radius for a circle centered at the reference point on the reference dose and tangent to the evaluation dose curve, but its sign is positive if the center of the circle is below the evaluation dose curve (e.g. Figure 2(a)) and negative if the center is above the curve (e.g. Figure 2(b)). Since the number of particles in the MC simulation is usually very large to ensure a small uncertainty level of clinical relevance; and with the large number of particles, the dose to a voxel in an MC simulation is commonly considered following a Gaussian distribution (Sempau and Bielajew, 2000). In this study, we assume that probability density function $p(t)$ is symmetric around zero. Note that $\tilde{\gamma} = \gamma_0 - t\cos\theta$ where $\theta$ is the angle between $\gamma_0$ and vertical axis, it is straightforward that

$$\gamma_0 = \int_{-\infty}^{+\infty} dt\, \tilde{\gamma}(t)\, p(t), \qquad (2)$$

For the average gamma index,

$$\bar{\gamma} = \int_{-\infty}^{+\infty} dt\, |\tilde{\gamma}(t)|\, p(t) = \int_{-\infty}^{\frac{\gamma_0}{\cos\theta}} dt\, \tilde{\gamma}(t)\, p(t) + \int_{\frac{\gamma_0}{\cos\theta}}^{+\infty} dt\, [-\tilde{\gamma}(t)]\, p(t). \qquad (3)$$

Now, subtracting Equations (2) and (3) leads to

$$\bar{\gamma} - \gamma_0 = 2 \int_{\frac{\gamma_0}{\cos\theta}}^{+\infty} dt\, [-\tilde{\gamma}(t)]\, p(t) \qquad (4)$$

Since $\tilde{\gamma}(t) < 0$ when $t > \frac{\gamma_0}{\cos\theta}$, we can conclude that

$$\bar{\gamma} \geq \gamma_0. \qquad (5)$$

This conclusion is valid for a general symmetric probability density function $p(t)$. It has also been theorized that the statistical fluctuations of the dose distribution in an MC calculation, termed as "noise" from here on, follow a Gaussian distribution (Sempau and Bielajew, 2000). So in this case, the average gamma index is then specified to be

$$\bar{\gamma} = \frac{1}{\sqrt{2\pi}\sigma_t} \int_{-\infty}^{+\infty} dt\, |\tilde{\gamma}(t)|\, e^{-t^2/2\sigma_t^2}, \qquad (6)$$

where $\sigma_t$ is statistical uncertainty value on the reference point. Then the Equation (6) can be rewritten as

$$\bar{\gamma} = \frac{1}{\sqrt{2\pi}\sigma_t} \int_{-\infty}^{\frac{\gamma_0}{\cos\theta}} dt\, (\gamma_0 - t\cos\theta) e^{-t^2/2\sigma_t^2}$$

$$+ \frac{1}{\sqrt{2\pi}\sigma_t} \int_{\frac{\gamma_0}{\cos\theta}}^{+\infty} dt\, (t\cos\theta - \gamma_0) e^{-t^2/2\sigma_t^2} \qquad (7)$$

$$= \gamma_0\, \mathrm{erf}\left(\frac{\gamma_0}{\sqrt{2}\sigma_t \cos\theta}\right) + \frac{2}{\sqrt{2\pi}} \sigma_t \cos\theta \exp\left(\frac{-\gamma_0^2}{2\cos^2\theta \sigma_t^2}\right). \qquad (8)$$

The derivative of Equation (8) with respect to $\sigma_t$ is

$$\frac{d\bar{\gamma}}{d\sigma_t} = \sqrt{\frac{2}{\pi}} \cos\theta \exp\left(\frac{-\gamma_0^2}{2\cos^2\theta \sigma_t^2}\right). \qquad (9)$$

Since $\frac{d\bar{\gamma}}{d\sigma_t} > 0$, and when $\sigma_t \to 0$, $\bar{\gamma} \to \gamma_0$, we get the same conclusion as Equation (5).

The increase of the average $\gamma$-index $\bar{\gamma}$ can be understood as following. When there is a finite deviation $t$, there are two scenarios resulting different impacts on the average $\gamma$-index. First, when $t$ is small for the original $\gamma$-index $\gamma_0$, i.e. $t \leq \frac{\gamma_0}{\cos\theta}$, the average of the $\gamma$-index pair $|\tilde{\gamma}(t)|$ and $|\tilde{\gamma}(-t)|$ equals $\gamma_0$. Second, when $t$ is large, i.e. $t > \frac{\gamma_0}{\cos\theta}$, the average of the pair $|\tilde{\gamma}(t)|$ and $|\tilde{\gamma}(-t)|$ is larger than $\gamma_0$ due to the flipped sign in one of them caused by the absolute value operation. It is the latter scenario that causes the increase of $\bar{\gamma}$. Hence when the original $\gamma$-index value is relatively large for the noise standard deviation, the increase of the average $\gamma$-index $\bar{\gamma}$ will be relatively small due to the small contributions from the second scenario.

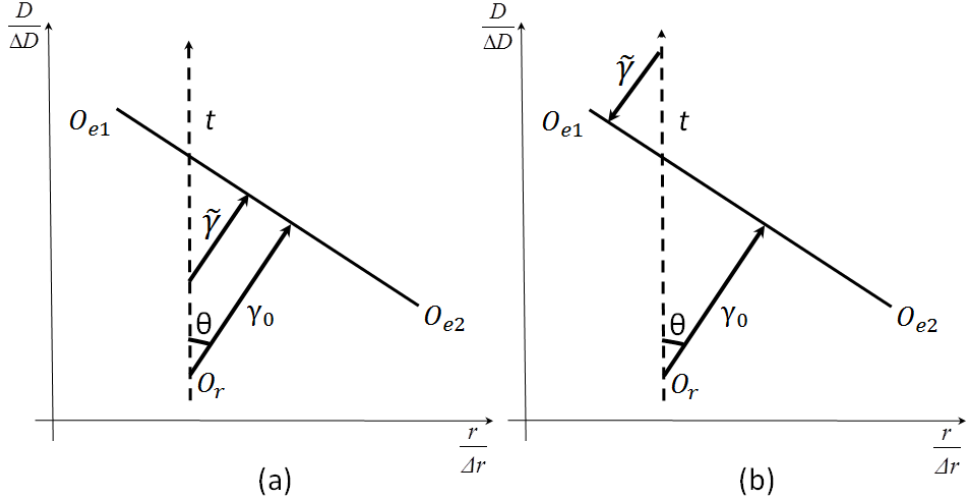

**Figures 2 (a) and (b).** Illustrations of two different contexts where noise is present in the reference dose.

*2.2.2 MC statistical fluctuations in the evaluation dose*

In Figures 3(a) and (b), suppose without noises, $O_{e1}O_{e2}$ is the line segment of the evaluation dose curve. With the MC noises, the line $O_{e1}O_{e2}$ moves. As in the Section 2.2.1, similarly we assume in this simplified model, the new γ-index value is only related to the line segment $O_{e1}O_{e2}$ of the evaluation dose curve, not to the other line segment of the evaluation dose curve. $\sigma_{t1}$, $\sigma_{t2}$ are the statistical uncertainties on the dose values at the points $O_{e1}$ and $O_{e2}$, the average γ-index value $\bar{\gamma}$ can be calculated as,

$$\bar{\gamma} = \int_{-\infty}^{+\infty} dt_1 \int_{-\infty}^{+\infty} dt_2\ |\tilde{\gamma}(t_1, t_2)| p(t_1, t_2), \tag{10}$$

where $t_i$, $i = 1, 2$, parameterizes the deviation of dose values at the two points $O_{e1}$ and $O_{e2}$ and $p(t_1, t_2)$ is the probability density function. For some noise realizations, the signed-gamma index $\tilde{\gamma}$ may change its sign from positive to negative. For instance, in Figure 3(a) when both $t_1$ and $t_2$ are large negative values. However, the probability of this situation is relatively small, given that $\sigma_{ti}$ are small for $\gamma_0$. Hence, we first ignore the contributions of γ-indices from these situations. The consequence when this contribution cannot be ignored will be discussed later. Given this assumption, all $\tilde{\gamma}$ are positive regardless of the values of $t_i$.

As for the probability distribution, for the large number of particles simulated in an MC dose calculation, we assume that $p(t_1, t_2) = p(-t_1, -t_2)$. When the noises at two points $O_{e1}$ and $O_{e2}$ are statistically independent, this assumption is apparently valid, as the noise distribution at each point is symmetric about zero. In reality, there exist correlations of noises between these two points. This correlation is caused by an electron track that passes through the two voxels. In one MC simulation, the number of electron tracks simultaneously passing the two voxels fluctuates about its average, and the probability of having more tracks is equal to that of having less tracks. Hence our assumption is still valid. Under this assumption, we can rewrite Equation (10) as

$$\bar{\gamma} = \frac{1}{2} \int_{-\infty}^{+\infty} dt_1 \int_{-\infty}^{+\infty} dt_2 [\tilde{\gamma}(t_1, t_2) + \tilde{\gamma}(-t_1, -t_2)] p(t_1, t_2). \tag{11}$$

From Figure 3(c), we further separate the integral domain into four different quadrants,

$$\bar{\gamma} = \frac{1}{2} \sum_{i=1}^{4} \iint_{I_i} dt_1 \, dt_2 [\tilde{\gamma}(t_1, t_2) + \tilde{\gamma}(-t_1, -t_2)] p(t_1, t_2). \tag{12}$$

Since the integrals in the domain $I_1$ and $I_2$ equal to those in the domain $I_3$ and $I_4$, respectively, Equation (12) reduces to

$$\bar{\gamma} = \sum_{i=1}^{2} \iint_{I_i} dt_1 \, dt_2 [\tilde{\gamma}(t_1, t_2) + \tilde{\gamma}(-t_1, -t_2)] p(t_1, t_2). \tag{13}$$

Figure 3(a) illustrates the context when $t_1$ and $t_2$ are both positive (in domain $I_1$). Suppose $O'_{e1} O'_{e2}$ and $\widetilde{O_{e1}O_{e2}}$ are the new evaluation dose lines, such that $O_{e1}\widetilde{O_{e1}}$ and $O_{e1}O'_{e1}$ are of the same length, similarly $O_{e2}\widetilde{O_{e2}}$ and $O_{e2}O'_{e2}$ are of the same length, and $\tilde{\gamma}(t_1, t_2)$ is represented as $O_r D$ while $\tilde{\gamma}(-t_1, -t_2)$ is $O_r A$. $O_r B$ is the original $\gamma$-index $\gamma_0$. If we set $CF \perp O'_{e1} O'_{e2}$, we have

$$O_r A + O_r D \leq O_r A + O_r C + CF. \tag{14}$$

Since $O_r A = O_r E \cdot \cos(\angle EO_r A)$ and $CF = CE \cdot \cos(\angle EO_r A)$, combining with Equation (14), we can get

$$O_r A + O_r D \leq O_r C \cdot [1 + \cos(\angle EO_r A)]. \tag{15}$$

Moreover $O_r B = O_r C \cdot \cos(\angle EO_r B)$. From the geometric relationship, we also have $\angle EO_r A = 180° - 2\alpha - 2\theta$ and $\angle EO_r B = 90° - \alpha - \theta$. Hence,

$$O_r A + O_r D \leq O_r C \cdot 2\cos^2(\angle EO_r B) \leq 2 \cdot O_r B. \tag{16}$$

This indicates that

$$\iint_{I_1} dt_1 \, dt_2 [\tilde{\gamma}(t_1, t_2) + \tilde{\gamma}(-t_1, -t_2)] p(t_1, t_2) \leq 2\gamma_0 \iint_{I_1} p(t_1, t_2) \, dt_1 \, dt_2. \tag{17}$$

Similarly, Figure 3(b) illustrates a context where $t_1$ is negative and $t_2$ is positive (in domain $I_2$). $\tilde{\gamma}(t_1, t_2)$ is represented as $O_r D$ while $\tilde{\gamma}(-t_1, -t_2)$ is $O_r A$. $O_r B$ is the original $\gamma$-index $\gamma_0$. With a similar derivation, we can generate the same conclusion as Equation (17).

$$\iint_{I_2} dt_1 \, dt_2 [\tilde{\gamma}(t_1, t_2) + \tilde{\gamma}(-t_1, -t_2)] p(t_1, t_2) \leq 2\gamma_0 \iint_{I_2} p(t_1, t_2) \, dt_1 \, dt_2. \tag{18}$$

Combine Equation (13), (17) and (18), we then have

$$\bar{\gamma} \leq 2\gamma_0 \left[ \sum_{i=1}^{2} \iint_{I_i} p(t_1, t_2) \, dt_1 \, dt_2 \right]. \tag{19}$$

Since $2 \sum_{i=1}^{2} \iint_{I_i} p(t_1, t_2) \, dt_1 \, dt_2 = 1$, we can conclude that

$$\bar{\gamma} \leq \gamma_0. \tag{20}$$

This indicates that the presence of noise will lead to an underestimation of the γ-index value. It is important to note that when the probability for negative signed γ-index $\tilde{\gamma}(t_1, t_2)$ is not negligible, i.e. when the noise standard deviation is relatively large for the original γ-index value, Equation (20) is no longer valid. The simplest case is when the reference dose distribution is the same as the evaluation dose and γ-index values are all zero. In this case, the MC noise in the evaluation dose distribution leads to an overestimation of γ-index values. However, as later shown in our numerical experiments, this situation is not clinically relevant because it only happens when the original γ-index values are very small and their variation due to the MC noise does not affect the γ-index passing rate.

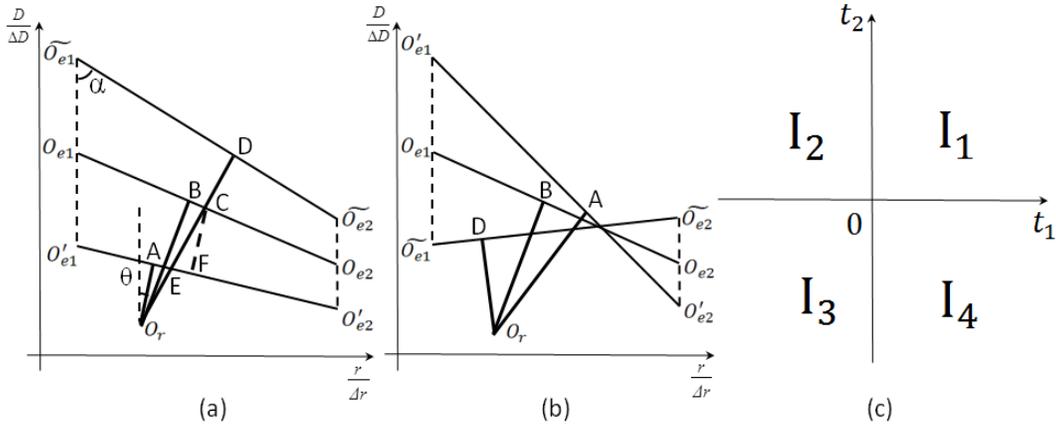

**Figures 3 (a) and (b).** Illustrations of two different contexts where noise is present in the evaluation dose. **(c).** Split of the integration domain for Equation (11).

*2.3 Numerical experiments*

We conducted numerical experiments on two realistic clinical cases for photon radiation therapy: a 7-beam IMRT prostate plan and a 2-arc VMAT HN plan. To study the effect of using an MC dose as the reference dose or the evaluation dose in γ-index tests, a non-MC dose distribution and a set of MC dose distributions at various $\sigma$ levels (including zero $\sigma$ level) are required for each clinical case. For the non-MC dose, we used the dose distribution in the patient plan extracted from a commercial treatment planning system (Eclipse, Varian Medical Inc., Palo Alto, CA) which was computed using the Analytical Anisotropic Algorithm (AAA). The resulting dose from the Eclipse system is at a resolution of 2.5×2.5×2.5 $mm^3$ and interpolated to the MC dose resolution of 1.953×1.953×2.5$mm^3$. The AAA algorithm is an analytical algorithm hence there is no noise in the dose. To get the set of MC dose distributions, we first extracted the MLC leaf sequences of all beam angles from the patient plan. Using different number of particle histories in the simulation, the dose distributions of various $\sigma$ level were calculated on the patient geometry using a GPU-based MC dose engine (gDPM) (Jia *et al.*, 2011). In this work we define the term $\sigma$ *level* as the average $\sigma$ value normalized to the maximum dose $D_{max}$ within regions of dose values higher than 50% of $D_{max}$ (VOI$_{50\%}$). The dose distributions with $\sigma$ level of 0.5%, 1%, 1.5%, and 2% were calculated, which are the most clinically relevant noise levels for MC dose calculations. An additional dose distribution of $\sigma$ level of 0.2% was also computed and used to obtain the MC dose with zero $\sigma$ level. As such we determined the de-noised dose value $D(\boldsymbol{v})$ by solving such an optimization problem

$$D(\boldsymbol{v}) = argmin_D E[D] = argmin_D \int d\boldsymbol{v} \left(D - \widehat{D} \log D\right) + \frac{\beta}{2} \int d\boldsymbol{v} \, |\nabla D|^2. \qquad (21)$$

The first term in the energy function $E[D]$ is a data-fidelity term considering the Poisson noise, while the second term is a penalty term to ensure the smoothness of the de-noised dose $D(\boldsymbol{v})$. Since the energy function is convex, the optimality condition is researched by solving

$$0 = \frac{\delta E}{\delta D} = \left(1 - \frac{\widehat{D}}{D}\right) - \beta \nabla^2 D. \tag{22}$$

This model is solved using a gradient descent method. We would like to point out that the beam parameters in the gDPM code were not purposely tuned to match the Eclipse results. To analyze the situation when both the evaluation and reference doses are generated by the MC method, another set of MC dose distributions at various $\sigma$ levels (including zero $\sigma$ level) is desired. For this set of MC doses, we used the same MC dose engine gDPM, but shifted the isocenter position of all beams by 3 mm to the patient's left, posterior, and superior directions, respectively, to simulate the dosimetric effects due to a set up error in a clinical situation.

To study the impact of the MC noise on the γ-index value, we first needed to conduct a base comparison between the non-MC dose and the MC dose with zero $\sigma$ level; the resulting γ-index values of a selected group of voxels are treated as the base values. Then the γ-index results for the same voxels were followed when the non-MC dose is compared with the MC doses of increasing $\sigma$ levels. These voxels are selected as the ones with more clinical relevance. Since the passing rate within a region of interest (ROI) is the most common criteria used to compare two dose distributions, we focused on a range of γ-index values that contribute to the calculation of the gamma passing rate. We also noticed that the behavior of average γ-index variation due to the MC noise is similar for close γ-index values. Thus, based on the base comparison, we selected four groups of voxels in the reference dose distribution with γ-index values from value 0.6 to 0.8 (0.6, 0.8), from value 0.8 to 1 (0.8, 1.0), from value 1 to 1.2 (1.0, 1.2), and from value 1.2 to 1.4 (1.2, 1.4). For each group of voxels, we followed the variation of the average γ-index value with the increase of the $\sigma$ level in the MC dose distribution. Furthermore, the γ-index passing rates were also reported for each comparison between the non-MC dose and the MC dose. In this study, in addition to $VOI_{50\%}$, we also selected another ROI where dose values were higher than 10% of $D_{max}$ ($VOI_{10\%}$).

Since the MC noise is a random variable, for the comparison at each $\sigma$ level, we repeated the γ-index test ten times for ten different random realizations of the dose distributions to get the mean of the γ-index test results. And during our experiments, all dose distribution comparisons were performed using the GPU-based fast γ-index algorithm (Gu *et al.*, 2011b).

## 3. Results

*3.1 MC reference dose vs Non-MC evaluation doses*

To study the effect of using an MC dose as the reference dose in γ-index tests, we treated the set of MC doses as the reference doses and the non-MC dose as the evaluation dose. Figures 4 (A) and (B) show the average γ-index values of voxels whose original γ-indices fall within the range of (0.6, 0.8), (0.8, 1.0), (1.0, 1.2) and (1.2, 1.4) as functions of the $\sigma$ level in the reference dose ($\overline{\sigma_{ref}}$) for both prostate and HN cases. For Figures 4(A) and (B), we used the most common clinical γ-index test criterion: 3%-3mm. We can see that the average γ-index value within each group slightly increases with $\overline{\sigma_{ref}}$. Figures 4(C) and (D) show the passing rate within VOI$_{50\%}$ and VOI$_{10\%}$ as function of $\overline{\sigma_{ref}}$ for two different γ-index test criteria: 3%-3mm and 2%-2mm. It is noted that, although the average γ-index value does not change much, the gamma passing rate decreases significantly with the increase of the σ levels in the reference dose, especially for VOI$_{50\%}$ for these two clinical cases.

　　　To better understand the effect of the MC noise on the gamma passing rate, we examined the voxels that contribute to the gamma passing rate calculation. We defined *Type-I voxels* as those with γ-index values larger than one in the base comparison when the MC σ level is zero and γ-index values smaller than or equal to one when the MC σ level is 2%. *Type-II voxels* are those with the opposite situation, *i.e.*, γ-index values increasing from below or equal to one to above one when the σ level increases from zero and 2%. Since the MC noise is a random variable, for a γ-index test with a random realization of the MC dose distribution, a particular voxel with the γ-index value around one can be either Type-I or Type-II. However, when running the γ-index test for many random realizations of the MC dose distributions, this voxel will have more chance to be Type-II than Type-I when using the MC dose as the reference dose. Table 1 summarizes the average percentages of Type-I and Type-II voxels within two ROIs for two comparison criteria after running the γ-index test for 10 different random realizations of the MC reference dose distributions. It is noted that percentage of Type-II voxels is higher than that of Type-I. The net effect, Type-II percentage minus the Type-I percentage, is equal to the change of the gamma passing rate with the MC dose of 2% σ level, as shown in Figures 4 (C) and (D).

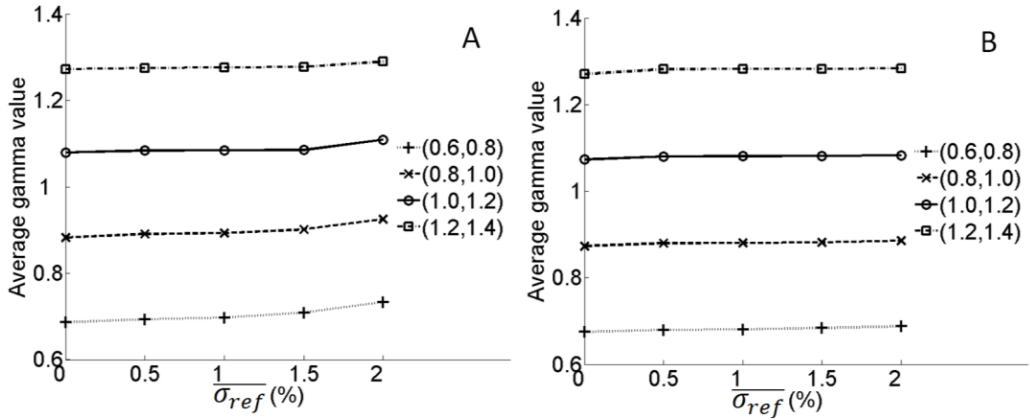

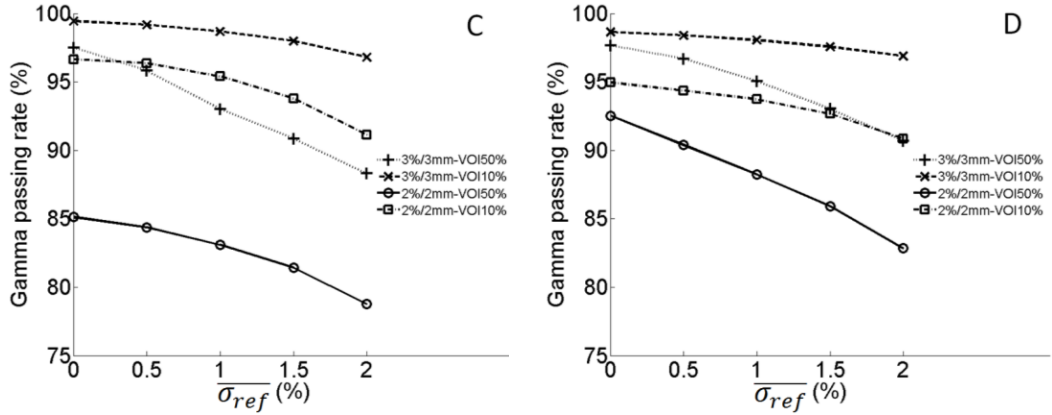

**Figure 4.** Average γ-index and gamma passing rate as functions of $\sigma$ level in the reference dose for MC reference doses vs non-MC evaluation dose. **(A) and (C):** prostate case; **(B) and (D):** HN case.

**Table 1.** Percentages of Type-I and Type-II voxels within $VOI_{50\%}$ and $VOI_{10\%}$ for MC reference doses of 2% $\sigma$ level vs the non-MC evaluation dose with 3%-3mm and 2%-2mm criteria.

| *Clinical case* | *Prostate* | | | | *HN* | | | |
|---|---|---|---|---|---|---|---|---|
| **Criteria** | **3%-3mm** | | **2%-2mm** | | **3%-3mm** | | **2%-2mm** | |
| **ROI** | $VOI_{50\%}$ | $VOI_{10\%}$ | $VOI_{50\%}$ | $VOI_{10\%}$ | $VOI_{50\%}$ | $VOI_{10\%}$ | $VOI_{50\%}$ | $VOI_{10\%}$ |
| Type-I Voxels (%) | 1.09 | 0.23 | 5.92 | 1.19 | 0.33 | 0.30 | 2.24 | 1.04 |
| Type-II Voxels (%) | 10.31 | 2.87 | 12.27 | 6.70 | 7.31 | 2.04 | 11.93 | 5.14 |
| Type-II − Type-I (%) | 9.22 | 2.64 | 6.35 | 5.51 | 6.98 | 1.74 | 9.69 | 4.10 |

*3.2 Non-MC reference dose vs MC evaluation doses*

To study the effect of MC noise in the evaluation dose in γ-index tests, we treated the non-MC dose as the reference dose and the set of MC doses as the evaluation doses. Figures 5 (A) and (B) show the average γ-index values of voxels whose original γ-indices fall within each range as functions of the $\sigma$ level in the evaluation dose ($\overline{\sigma_{eva}}$) for prostate and HN cases. The average γ-index value within each group decreases dramatically with $\overline{\sigma_{eva}}$. Figures 5 (C) and (D) show, for two different γ-index test criteria, 3%-3mm and 2%-2mm, the passing rate within $VOI_{50\%}$ and $VOI_{10\%}$ as function of $\overline{\sigma_{eva}}$. We observed that, the gamma passing rates saturated for 3%-3mm criterion, while the gamma passing rates for 2%-2mm criterion increases with the increase of $\overline{\sigma_{eva}}$, especially for $VOI_{50\%}$. Table 2 summarizes the average percentage of Type-I or Type-II voxels within the ROIs. For the non-MC reference dose vs the MC evaluation dose of 2% $\sigma$ level, the percentage of Type-I voxels is higher than that of Type-II which means that, statistically, there are more voxels where the γ-index value sinks below one than voxels where the γ-index value rises above one. The net difference between Type-I percentage and Type–II percentage, is the same as the change of the gamma passing rate with the MC dose of 2% $\sigma$ level, shown in Figures 5 (C) and (D).

In Section 2.2.2, we have noticed that, when the statistical standard deviation in the MC evaluation dose distribution is relatively large for the original γ-index value, the γ-index value will be overestimated on average due to the noise. Figures 6 (A) and (B) show the average γ-index value of voxels with small original γ-index values, ranging from 0.1 to 0.24, as functions of $\overline{\sigma_{eva}}$ for the prostate and HN cases. From Figure 6 we can see that when the original γ-index value is large enough, i.e. larger than 0.2, the average γ-index decreases with $\overline{\sigma_{eva}}$ (shown in blue and dashed lines); however, when the original γ-index value is very small, i.e. smaller than 0.14, the average γ-index increases with $\overline{\sigma_{eva}}$ (shown in green and dotted lines). For the in-between original γ-index value, the average γ-index first decreases and then increases with $\overline{\sigma_{eva}}$ (shown in red and dashed/dotted lines). We would like to point out that voxels with this range of original γ-index values do not contribute to the passing rate change.

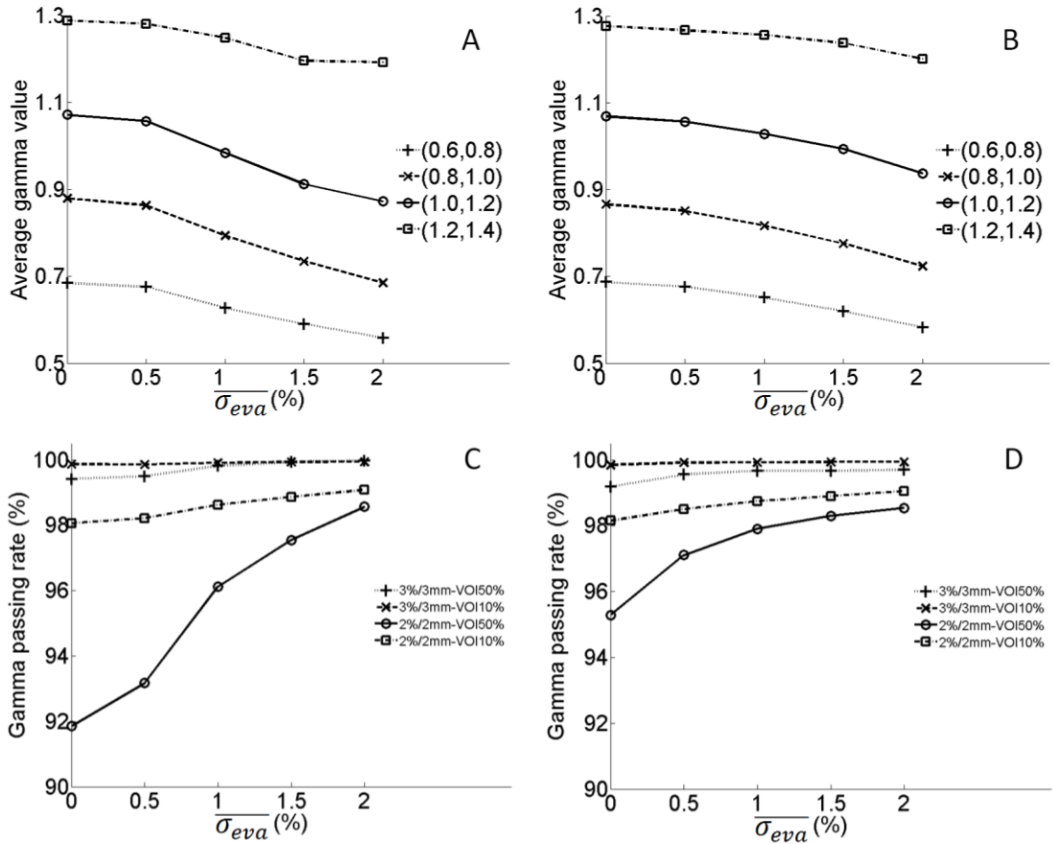

**Figure 5.** Average γ-index and gamma passing rate as functions of *σ* level in the evaluation dose for non-MC reference dose vs MC evaluation doses. **(A) and (C):** prostate case; **(B) and (D):** HN case.

**Table 2.** Percentages of Type-I and Type-II voxels within VOI50% and VOI10% for the non-MC reference dose vs MC evaluation doses of 2% *σ* level with 3%-3mm and 2%-2mm criteria.

| *Clinical case* | *Prostate* | | | | *HN* | | | |
|---|---|---|---|---|---|---|---|---|
| **Criteria** | **3%-3mm** | | **2%-2mm** | | **3%-3mm** | | **2%-2mm** | |
| **ROI** | $VOI_{50\%}$ | $VOI_{10\%}$ | $VOI_{50\%}$ | $VOI_{10\%}$ | $VOI_{50\%}$ | $VOI_{10\%}$ | $VOI_{50\%}$ | $VOI_{10\%}$ |
| Type-I Voxels (%) | 0.50 | 0.09 | 7.20 | 1.22 | 0.510 | 0.097 | 3.63 | 1.06 |
| Type-II Voxels (%) | 0.02 | 0.01 | 0.48 | 0.20 | 0.004 | 0.003 | 0.38 | 0.16 |
| Type-I – Type-II (%) | 0.48 | 0.08 | 6.72 | 1.02 | 0.506 | 0.094 | 3.25 | 0.90 |

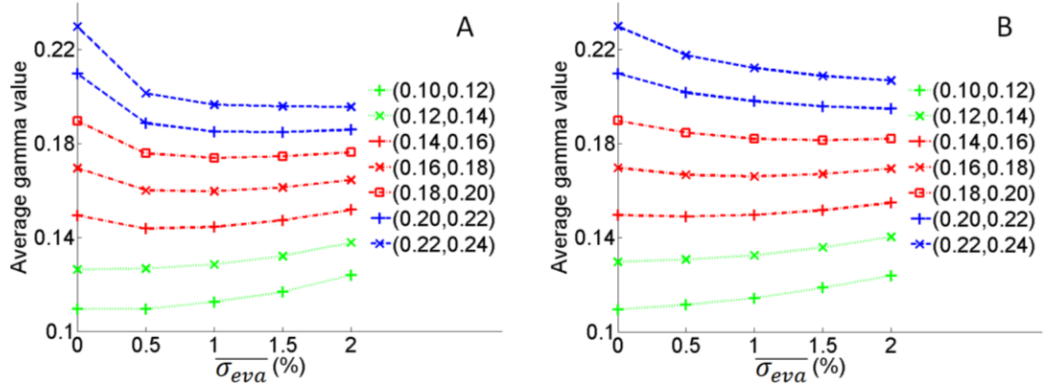

**Figure 6.** Average γ-index as functions of $\sigma$ level in evaluation dose for non-MC reference dose vs MC evaluation doses. **(A):** prostate case; **(B):** HN case.

*3.3 MC reference dose vs MC evaluation dose*

To analyze the situation when both the evaluation and reference doses are generated by the MC method, we considered the set of MC doses as the reference doses and the other set of MC doses with the shifted isocenter as the evaluation doses. Figure 7 shows the color maps of the gamma passing rate in high dose region $VOI_{50\%}$ under 3%-3mm criterion. The $x$-axis is the $\sigma$ level in the evaluation dose, while the $y$-axis is the $\sigma$ level in the reference dose. The values at origin of the two maps are the base value with zero $\sigma$ level in both reference and evaluation doses. Along the $x$-axis, the results correspond to the cases for the non-MC reference dose versus MC evaluation doses, same as in Figures 5 (C) and (D). Along the $y$-axis, the results correspond to the cases for MC reference doses versus the non-MC evaluation dose, same as in Figures 4 (C) and (D). The black lines in Figure 7 illustrate iso-value lines on which the passing rate is the same as the base value for MC doses of zero $\sigma$ level. The iso-value line splits the map into two regions: the upper-left region, where the MC noise level is relatively high in the reference dose leading to the underestimation of the gamma passing rate, and the lower-right region, where the MC noise level is relatively high in the evaluation dose leading to the overestimation of the gamma passing rate. The shape of iso-value line and the way that it splits the map are case-dependent. When both doses are the MC doses, we redefine *Type-I voxels* as those with γ-index values larger than one when the $\sigma$ level is zero in both the evaluation and the reference doses and less than or equal to one when the $\sigma$ level is 2% in both doses; *Type-II voxels* are those with the opposite situation, *i.e.*, γ-index values increasing from below or equal to one to above one when both $\sigma$ levels increase from zero and 2%. From Table 3, the average percentage of Type-I is higher than the percentage of Type-II voxels and the net contribution matches with the increased gamma passing rate in the right upper corner with 2% $\sigma$ level in both reference and evaluation doses.

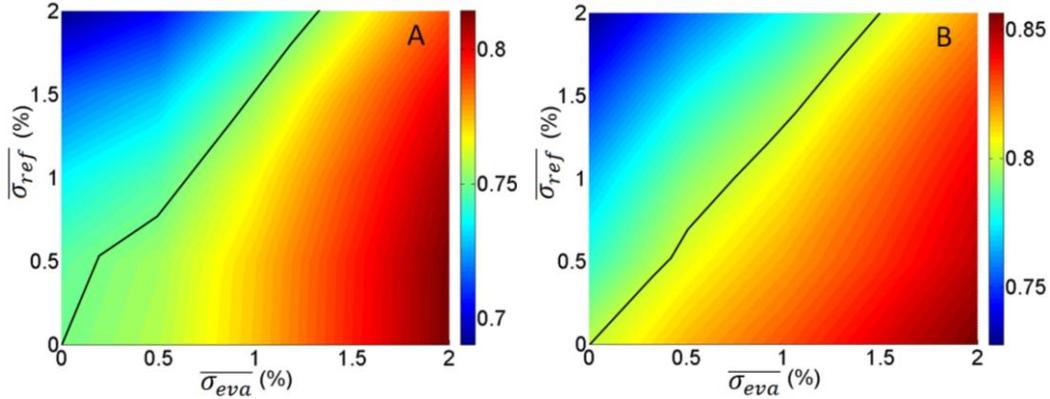

**Figure 7.** The color maps of gamma passing rate within VOI50% as functions of *σ* level in the reference and evaluation doses for MC reference dose vs MC evaluation dose with 3%-3mm criterion. **(A):** prostate case; **(B):** HN case.

**Table 3.** Percentages of Type-I and Type-II voxels within VOI50% for the MC reference dose of 2% *σ* level vs the MC evaluation dose of 2% *σ* level with 3%-3mm criterion.

| Clinical case | Type-I Voxels (%) | Type-II Voxels (%) | Type-I – Type-II (%) |
|---|---|---|---|
| *Prostate* | 7.80 | 3.91 | 3.89 |
| *HN* | 6.54 | 4.39 | 2.15 |

## 4. Discussion and Conclusions

In this paper, we have first demonstrated in a simplified 1D model that, to the first order approximation, MC statistical fluctuation in the reference dose tends to overestimate the γ-index, while that in the evaluation dose tends to underestimate the γ-index when the original γ-index value is relatively large. This simplified model does not serve as a strict theoretical proof but rather as a theoretical guidance for 2D and 3D cases. To validate the theoretical conclusions, we conducted numerical experiments on two clinical cases for photon radiation therapy: an IMRT prostate case and a VMAT HN case. We focused on voxels with clinically relevant γ-index values in the absence of noise, range of 0.6 to 1.4, and found that when performing γ-index tests between an MC reference dose and a non-MC evaluation dose, the average γ-index is overestimated but the change is not significant. Second, when performing γ-index tests between a non-MC reference dose and an MC evaluation dose, the average γ-index is underestimated and decreases with the increase of noise level in the evaluation dose. This is doubly confirmed by the blue dashed curves in the Figure 6. When the original γ-index value $\gamma_0$ is larger than 0.2, the average γ-index monotonically decreases with the increase of the noise level. For the green dotted curves in the Figure 6, when the $\gamma_0$ is smaller than 0.14, the average γ-index increases with the noise level. For those cases with $\gamma_0$ lying in between the above two limits, the average γ-index first decreases when the noise level is low and then increases. Nonetheless, in the latter two situations, the changes of γ-index values due to the MC noise are not expected to considerably impact the gamma passing rates, since the small original γ-index values are much smaller than one. Hence these two situations are not clinically relevant.

The change for the gamma passing rate within the ROI due to the MC noise is most relevant for the clinical applications of γ-index test. In the experiment, we defined two quantities, percentage of Type-I voxels and percentage of Type-II voxels and we found the following. 1) When performing γ-index test between an MC reference dose and a non-MC evaluation dose, the gamma passing rate decreases with the increase of the statistical noise level in the reference dose. 2) When performing γ-index test between a non-MC reference dose and an MC evaluation dose, the gamma passing rate increases with the increase of the noise level in the evaluation dose. In these two situations, the magnitude of the change of gamma passing rate when 2% $\sigma$ level exists in the MC dose equals to the difference between the percentage of Type-I and that of Type-II voxels. 3) When the reference dose and the evaluation dose are both MC doses, the gamma passing rate increases when the statistical noise in the evaluation dose increases. It decreases when the statistical noise in the reference dose increases. Considering again the correlation between the neighboring voxels in the MC evaluation dose, this effect on the γ-index is usually local in the spatial domain. However, since the gamma passing rate is a statistical overall effect from all the voxels within the ROI, the local effect will be smeared out in the whole ROI.

For the two clinical cases we have tested, the effect on the gamma passing rate is quite significant. Taking the 3%-3mm test criterion as an example, when there exists 2% MC $\sigma$ level in the reference dose, the gamma passing rate in $VOI_{50\%}$ drops from 97.5% to 88.3% and in $VOI_{10\%}$ from 99.4% to 96.8% for the prostate case. For the HN case, the gamma passing rate in $VOI_{50\%}$ drops from 97.7% to 90.7% and in $VOI_{10\%}$ from 98.6% to 96.9%, respectively. On the other hand, when 2% MC noise level exists in the evaluation dose, the resulting increase of the gamma passing rate is not significant under the 3%-3mm criterion. This is because the passing rate is already very close to one in the absence of noise for this relatively loose criterion. However, the changes are more obvious under 2%-2mm criterion. Especially in $VOI_{50\%}$, it increases from 91.9% to 98.6% for the prostate case and from 95.3% to 98.5% for the HN case. Based on our theoretical and numerical results, we conclude that great caution is needed when dealing with MC doses in γ-index tests. The MC statistical fluctuation effect should be considered when analyzing the γ-index test results to avoid biased conclusions. In practice, we should try to alleviate this problem. A straightforward approach is to simulate a large number of particle histories in an MC dose calculation to reduce the MC statistical uncertainty. Additionally, denoising the MC dose results can also be performed.

The conclusions based on our theoretical analysis were verified using Monte Carlo simulations with clinical photon beams. The studies for other types of clinical radiation beams such as electron, proton, and heavy ion beams will be performed in the future.

**Acknowledgement**

This work is supported in part by a Master Research Agreement from Varian Medical Systems, Inc.

**References**

Bakai A, Alber M and Nusslin F 2003 A revision of the gamma-evaluation concept for the comparison of dose distributions *Physics in Medicine and Biology* **48**

Calvo O I, Gutierrez A N, Stathakis S, Esquivel C and Papanikolaou N 2012 On the quantification of the dosimetric accuracy of collapsed cone convolution superposition (CCCS) algorithm for small lung volumes using IMRT *Journal of Applied Clinical Medical Physics* **13**

Chen M, Lu W, Chen Q, Ruchala K and Olivera G 2009 Efficient gamma index calculation using fast Euclidean distance transform *Physics in Medicine and Biology* **54**

Cheng A, Harms W B, Gerber R L, Wong J W and Purdy J A 1996 Systematic verification of a three-dimensional electron beam dose calculation algorithm *Medical Physics* **23** 685-93

Depuydt T, Van Esch A and Huyskens D P 2002 A quantitative evaluation of IMRT dose distributions: refinement and clinical assessment of the gamma evaluation *Radiotherapy and Oncology* **62**

Gu X, Jelen U, Li J, Jia X and Jiang S B 2011a A GPU-based finite-size pencil beam algorithm with 3D-density correction for radiotherapy dose calculation *Physics in Medicine and Biology* **56** 3337-50

Gu X, Jia X and Jiang S B 2011b GPU-based fast gamma index calculation *Physics in Medicine and Biology* **56** 1431

Harms W B, Low D A, Wong J W and Purdy J A 1998 A software toes for the quantitative evaluation of 3D dose calculation algorithms *Medical Physics* **25** 1830-6

Hissoiny S, Ozell B, Bouchard H and Despres P 2011 GPUMCD: A new GPU-oriented Monte Carlo dose calculation platform *Medical Physics* **38** 754-64

Jelen U and Alber M 2007 A finite size pencil beam algorithm for IMRT dose optimization: density corrections *Physics in Medicine and Biology* **52** 617-33

Jia X, Gu X, Graves Y J, Folkerts M and Jiang S B 2011 GPU-based fast Monte Carlo simulation for radiotherapy dose calculation *Physics in Medicine and Biology* **56** 7017-31

Jia X, Gu X, Sempau J, Choi D, Majumdar A and Jiang S B 2010 Development of a GPU-based Monte Carlo dose calculation code for coupled electron-photon transport *Physics in Medicine and Biology* **55**

Jiang S B, Sharp G C, Neicu T, Berbeco R I, Flampouri S and Bortfeld T 2006 On dose distribution comparison *Physics in Medicine and Biology* **51** 759-76

Ju T, Simpson T, Deasy J O and Low D A 2008 Geometric interpretation of the gamma dose distribution comparison technique: Interpolation-free calculation *Medical Physics* **35** 879-87

Low D A 2010 Gamma dose Distribution Evaluation Tool *Journal of Physics: Conference Series* **250**

Low D A and Dempsey J F 2003 Evaluation of the gamma dose distribution comparison method *Medical Physics* **30** 2455-64

Low D A, Harms W B, Mutic S and Purdy J A 1998 A technique for the quantitative evaluation of dose distributions *Medical Physics* **25** 656-61

Sempau J and Bielajew A F 2000 Towards the elimination of Monte Carlo statistical fluctuation from dose volume histograms for radiotherapy treatment planning *Physics in Medicine and Biology* **45** 131-57

Shiu A S, Tung S, Hogstrom K R, Wong J W, Gerber R L, Harms W B, Purdy J A, Haken R K T, McShan D L and Fraass B A 1992 VERIFICATION DATA FOR ELECTRON-BEAM DOSE ALGORITHMS *Medical Physics* **19**

Stock M, Kroupa B and Georg D 2005 Interpretation and evaluation of the gamma index and the gamma index angle for the verification of IMRT hybrid plans *Physics in Medicine and Biology* **50**

Vandyk J, Barnett R B, Cygler J E and Shragge P C 1993 COMMISSIONING AND QUALITY ASSURANCE OF TREATMENT PLANNING COMPUTERS *International Journal of Radiation Oncology Biology Physics* **26** 261-73

Wang H, Ma Y, Pratx G and Xing L 2011 Toward real-time Monte Carlo simulation using a commercial cloud computing infrastructure *Physics in Medicine and Biology* **56** N175-N81

Wendling M, Zijp L J, McDermott L N, Smit E J, Sonke J-J, Mijnheer B J and van Herk M 2007 A fast algorithm for gamma evaluation in 3D *Medical Physics* **34**

Yuan J and Chen W 2010 A gamma dose distribution evaluation technique using the k-d tree for nearest neighbor searching *Medical Physics* **37**